\def\beg{\begin{equation}}
\def\eeq{\end{equation}}
\documentstyle[12pt]{article}
\begin{document}
\begin{center}
{\Large{\bf Quasiparticles in quantum Hall effect: Smet's  fractional
 charges.}}
\vskip0.35cm
{\bf Keshav N. Shrivastava}\\
{\it School of Physics, University of Hyderabad,
Hyderabad 500046, India}
\end{center}

It has been pointed out by Smet that there are fractional-charge 
values which do not fit their formula of composite fermions. We find 
that our formula predicts these fractional charges very well and in 
fact there exists a relationship between spin and the effective 
charge of a quasiparticle.

\vskip0.10cm
Corresponding author: keshav@mailaps.org\\
Fax: +91-40-2301 0145.Phone: 2301 0811.
\vskip0.10cm

\noindent {\bf 1.~ Introduction}

    For some years it was thought that ``flux quanta" are attached to 
the electron. Considerable effort was put into this ``flux attached" 
model to explain the quantum Hall effect. An ``incompressible" model 
was suggested but it turns out that $a_o$=1 is not a solution of the
 algebraic equations involved, $Ba_o^2$=$\phi_o$. Here $a_o$=1 is 
required for incompressibility. After a lapse of twenty years, now Pan 
et al[1] have pointed out that quantum Hall effect occurs at some of 
the fractions which can not be obtained by their model. It was thought
 that plateaus occur at the filling factors, $\nu$=$p/(2mp\pm 1)$, 
where $m$ and $p$ are integers. Now, the recently reported fractions
 do not 
fit with this formula. Smet[2] has also agreed that the formula, 
$\nu$=$p/(2mp\pm 1)$ of the composite fermion model (CF) has no 
rigorous foundation. Similarly, composite bosons (CB) will not be 
feasible. The flux quanta are not so abundant, such as 10 per 
electron, that they can attach to electrons. Therefore, the idea 
of `` flux quanta" attachment to the electrons should be discarded. 
Considerable effort has been made to examine many papers in which 
a claim is made that flux quanta are attached to the electrons but
 in all cases 
internal inconsistencies have been found.  References to some of 
the papers which point out the inconsistencies are as follows.

\noindent cond-mat/0209666 shows that CF violates classical 
electrodynamics;\\
0210320 shows that there are far too many parameters in CF model;\\
0211223 shows that CF effective field is incorrect; \\
0301380 shows that CF requires E, H decoupling in Maxwell equations;\\
0302009 shows that CF and CB transformation has not been carried out;\\
0302315 shows that mass of the CF is not consistent with the available
 space;\\
0302461 shows what quantity is measured in fractional charge
 experiments;\\
0303146 shows that CF lacks in Lorentz invariance;\\
0303014 shows that CF model is internally inconsistent.\\

     Usually a Jordan-Wigner transformation transforms the spin 
operators into fermions. Similarly, the Holstein-Primakopf transformation 
transforms the spin operators into magnons. Under 
special conditions, such as zero temperature in the case of Jordan-Wigner
 and low temperatures in the case of Holstein-Primakopf, the solution of 
the transformed hamiltonian is quite close to that of the starting
 hamiltonian so that some useful results can be obtained. However, in 
the
case of ``flux attachment" no transformation has been found. Therefore,
 there is no way except to discard the ``CF" model[3]. Dyakonov[4] has 
clearly noted that CF model is not based on good theoretical principles
 and Farid[5] has pointed out that the field formula is not correct.
We are then left with no solution of the problem of quantum Hall effect.
We have found the correct theory of the quantum Hall effect[6].
 According to this theory all of the fractional charges which do not 
fit in the CF model, are correctly predicted. In particular, the 
fractions, observed by Pan et al[1] have been well predicted[7]. At an
 earlier time, Pan et al[8] thought that even feature is present in 
the data. However, we have shown[9] that the experimental data is not
 consistent with the even feature. Here even feature means the quantity
 $2mp$ which occurs in the denominator.

      In this paper, we show that the fractional charges observed by 
Smet[2] are well predicted by our theory.

\noindent{\bf 2.~~Smet's observation}

Motivated by the work of Pan et al[1], Smet[2] searched for the
 fractional charges which are not found in the composite fermion model.
Smet searched the region in the center of 2/5 and 1/3. Small steps 
were found at,
\beg
\nu= 4/11, 7/19, 10/27, ..., 11/29, 8/21, 5/13
\eeq
in the experimental measurement of transverse resistivity. These
 fractions do not fit the CF model. However, if 1/2 of these values
 could fit the formula, it will be sufficient and we could declare 
that these are two-particle states but for integer values of $m$ and
 $p$ the formula $p/(2mp\pm 1)$ does not fit the data. For two
 particle bound state, there will be the need for a binding energy 
and it becomes more difficult to fit the CF model with the experimental 
data.

     W\"ojs et al have suggested to use $2{\it l}+1$ as the denominator
 as in our theory and add an arbitrary pseudopotential parameter untill
 the CF agrees with the data, if not for one particle, then for any 
two-particle bound state. This is obviously arbitrary and not like physics because of the  arbitrary parameter. Therefore, W\"ojs et al have not solved the problem. There is a suggestion that the fractional charges  like 4/11, etc. are generated by a process similar to fractals. The fractal model of phase transition with fractional dimensionality does not satisfy the quantum Hall effect data. A nuclear decay type model also does not satisfy the data.

\noindent{\bf4.~~Theory}

     Our theory is explained in a recent book[11]. Needless to say 
that
this theory agrees with the quantum Hall effect data. We find that
 Laughlin has not resolved whether the area becomes fractional or 
the charge. Since the magnetic area is $a_o^2$=$\phi_o/B$, it is 
important to know that the quantity which enters in the Laughlin's 
paper is the product $eB$ and not $e$ alone. Schrieffer has used 
the Laughlin's work as if it is $e$. There is no serious problem 
created by this type of treatment as long as we keep track of the 
product $eB$. We have explained our theory in comparison with Laughlin
 and Schrieffer in ref.[12]. Ref.[7] shows the interpretation of the 
data of Pan et al[1].
At this stage our attention was drawn to a paper by Mani and von
 Klitzing[13] which has 146 different fractions. Surprizingly, we are 
able to understand all of these 146 values [14,15]. Pan et al[1] are
 concerned with 4/11, 5/13, 6/17, 4/13, 5/17 and 7/11. Let us add 
these 6 values to our 146 values so that it may be thought that 152
 values come out to be correct from our theory. To this we add another
 6 values from Smet[2] so that we have 158 values. The interpretation 
of Smet's values is given below. In our theory, the effective charge
 is given by the formula,
\beg
e_{eff}/e={{\it l}+{1\over 2}\pm s\over 2{\it l}+1}
\eeq
which gives 4/11 for ${\it l}$=5 and $s$=-3/2, 7/19 comes for
 ${\it l}$=9, $s$=-5/2; 10/27 comes for ${\it l}$=13, $s$=-7/2; 
11/29 comes for ${\it l}$=14, $s$=-7/2; 8/21 has ${\it l}$=10, 
$s$=-5/2; 5/13 has ${\it l}$=6 and $s$=-3/2. Thus 4/11 belongs to a 
cluster of 3 electrons with ${\it l}$=5 and all polarized with negative
 sign for the spin. The fraction 7/19 has five electrons in
 ${\it l}$=9 with all spin polarized with negative sign, etc. Thus all
 of Smet's experimentally measured values fit well in our formula 
showing that there are clusters of electrons with a small number of 
electrons in each cluster.  For ${\it l}$=7, $s$=-3/2, ${\nu_{-}}$=6/15;
 for ${\it l}$=7, $s$=-1/2, $\nu_{-}$=7/15; for ${\it l}$=8, $s$=-3/2,
 $\nu_{-}$=7/17, etc are predicted but not noted by Smet from 
experimental work. Thus we learn that there is a linear relationship
 between charge and spin. For ${\it l}$=0, the charge $\nu_{\pm}$ is 
related to spin, $s$ as,
\beg
\nu_{\pm}={1\over 2}\pm s.
\eeq
We can tabulate this expression as in Table 1 below.
\begin{center}
{\bf Table 1}: Predicted fractional charge for ${\it l}$=0\\
\begin{tabular}{cccc}
\hline
S.No. &  $s$ & $\nu_+$ & $\nu_-$\\
\hline
1     &  1/2 & 1       &   0\\
2     &  3/2 & 2       &  -1\\
3     &  5/2 & 3       &  -2\\
4     &  7/2 & 4       &  -3\\
\hline
\end{tabular}
\end{center}

This means that in the case of spherically symmetric states, half
 integer spin gives rise to integer charge with pairwise production 
of quasiparticles such that $\nu_++\nu_-$=1 but there is spin 
polarization.
The zero charge for $s$=1/2 means that there is a charge-density 
wave which is linked to distortions in the solid. Integer charge is 
seen in electron clusters. For ${\it l}$=0, one electron gives rise
 to quasiparticles of zero charge and also equal to that of one
 electron. Three electrons must occupy three sites  and then the
 quasiparticle charge may be 2 for one quasiparticle and -1 for the
 other, etc. Why should three electrons give a charge of 2? Actually, 
it is quite 
simple to understand. If all the three electrons are at one point,
 then it is justified that charge should be 3, otherwise, charge must 
be 
transported to one point before it is added. Fractional charge is 
developed due to this transport.

\noindent{\bf3.~~ Conclusions}.

     It is clear that our formula predicts all of Smet's values correctly. We have explained[16] that composite fermion(CF) model is internally inconsistent and the transformation from electrons to composite fermions has not been done. In any case, the composite 
fermion series $p/(2mp\pm 1)$ is inconsistent with the experimental 
data of Smet. Our theory[11, 12, 14,15] is based on quantum mechanics 
and works very well on all of the experimentally observed values of the
 fractional charges. We have discovered  a new spin-charge relationship. 

\vskip0.25cm

\noindent{\bf4.~~References}
\begin{enumerate}
\item W. Pan, H. L. Stormer, D. C. Tsui, L. N. Pfeiffer, K. W. Baldwin 
and K. W. West, Phys. Rev. Lett. {\bf 90}, 016801(2003).
\item J. H. Smet, Nature {\bf 422}, 391(2003).
\item K. N. Shrivastava, Bull. Am. Phys. Soc. J1.209(2003).
\item M. I. Dyakonov, cond-mat/0209206.
\item B. Farid, cond-mat/0003064.
\item K. N. Shrivastava, Bull. Am. Phys. Soc. J1.108(2003).
\item K. N. Shrivastava, cond-mat/0302610[Interpretation of ref.1]
\item W. Pan, et al, Phys. Rev. Lett.{\bf 88},176802 (2002)
\item K. N. Shrivastava, cond-mat/0204627.
\item A. W\"ojs, K-S. Yi and J. J. Quinn, cond-mat/0304130.
\item K. N. Shrivastava, Introduction to quantum Hall effect,\\ 
      Nova Science Pub. Inc., N. Y. (2002).
\item K. N. Shrivastava, cond-mat/0212552.
\item R. Mani and K. von Klitzing, Z. Phys. B{\bf 100}, 635 (1996).
\item K. N. Shrivastava, cond-mat/0303309.
\item K. N. Shrivastava, cond-mat/0303621.
\item K. N. Shrivastava, cond-mat/0304014.
\end{enumerate}
\vskip0.1cm

Note: Ref.11 is available from:\\
 Nova Science Publishers, Inc.,
400 Oser Avenue, Suite 1600,\\
 Hauppauge, N. Y.. 11788-3619,
Tel.(631)-231-7269, Fax: (631)-231-8175,\\
 ISBN 1-59033-419-1 US$\$69$.
E-mail: novascience@Earthlink.net

\end{document}